\documentclass[conference]{IEEEtran}
\ifCLASSINFOpdf
  \usepackage[pdftex]{graphicx}
 \DeclareGraphicsExtensions{.pdf,.jpeg,.png}
\else
\fi
\begin{document}
%
\title{A.Q.M.E.I.S.: Air Quality Meteorological and Enviromental Information System in Western Macedonia, Hellas.}

\author{\IEEEauthorblockN{Ioannis A. Skordas}
\IEEEauthorblockA{Lab. of Atmospheric Pollution \\ \& Environmental Physics \\ Dept. of Geotechnology \\
and Environmental Engineering\\ Technological Educational\\ Inst. of Western Macedonia,\\ Kozani, Hellas}
\and
\IEEEauthorblockN{George F. Fragulis}
\IEEEauthorblockA{Lab. of Web Technologies \\  \& Applied Control Systems\\ Dept. Of Electrical Engineering\\
Technological Educational \\Inst. of Western Macedonia, \\Kozani, Hellas}
\and
\IEEEauthorblockN{Athanassios G. Triantafyllou}
\IEEEauthorblockA{Lab. of Atmospheric Pollution \\ \& Environmental Physics \\ Dept. of Geotechnology \\and Environmental Engineering\\
Technological Educational \\Inst. of Western Macedonia, \\Kozani, Hellas}}


%


\maketitle

\begin{abstract}
An operational monitoring, as well as high resolution local-scale meteorological and air quality forecasting information system for Western Macedonia, Hellas, has been developed and is operated by the Laboratory of Atmospheric Pollution and Environmental Physics / TEI Western Macedonia since 2002, continuously improved. In this paper the novelty of information system  is presented, in a dynamic, easily accessible and user-friendly manner.
It consists of a structured system that users have access to and they can manipulate thoroughly, as well as of a system for accessing and managing results of measurements in a direct and dynamic way. It provides updates about the weather and pollution forecast for the next few days (based on current day information) in Western Macedonia. These forecasts are displayed through dynamic-interactive web charts and the visual illustration of the atmospheric pollution of the region in a map using images and animation images. 
\end{abstract}


%
\IEEEpeerreviewmaketitle

\section{Introduction}
 Web applications constitute valuable up-to-date tools in a number of different cases. One such case is their use in the management of environmental problems so as to protect civilians from any unfortunate consequences that these problems can cause. Their evolution, therefore, has been especially important in many cases, one of them being in the development of systems of administration of the air quality \cite{Triant2004}.The right of accessing  environmental information has been  enacted in European level through appropriate legislation, which are incorporated in the relevant Greek legislation see \cite {Council1}- \cite {Council5}.

Nowadays, the combination of telecommunications and new technologies create a framework for developing such systems increasingly sophisticated \cite {Karatzas}- \cite {Triant_book} . That is just diffusion of environmental information and public access which was attempted effectively through the system codenamed EAP (Laboratory of Atmospheric Pollution and Environmental Physics) in Western Macedonia. It is  developed for the first time in 2002  \cite {anakoinosi}, providing the possibility for direct information to the public about the air quality, as it was recorded in the four atmospheric measurement stations established in the capitals of Countries Kozani, Florina, Kastoria and Grevena though an appropriate web-site, as well as SMS, with the possibility for extension of stations and also the historical measurements privilege \cite {triantEvazoras2006}. For every station a previous and current index of pollution appears (in a scale 1-10) with an appropriate colour scale \cite {Comeap}.
The system was expanded and upgraded in May 2010, which consists in transferring data, the way of presentation as well as the amount of information provided. Specifically it is recommended: a) the combine use of different methods of transportation in real or almost real time data of terminal stations measurements to a central base station b) the environmental information is promoted to the internet, with a properly designed dynamic website with enabled navigating of Google map  \cite {TriantSkordas}, \cite {Skordas_Fragulis_Triant2011}, \cite {airlab}.

In this paper the novelty of information system EAP is presented, in a dynamic, easily accessible and user-friendly manner. It consists of a structured system that users have access to and they can manipulate thoroughly, as well as of a system for accessing and managing results of measurements in a direct and dynamic way. It provides updates about the weather and pollution forecast for the next few days (based on current day information) in Western Macedonia. These forecasts are displayed through dynamic-interactive web charts and the visual illustration of the atmospheric pollution of the region in a map using images and animation images. Moreover, there is the option to view historical elements. An additional new function is the use of online reports to monitor, analyze, control and processing measurements, historical data and statistics of each station in real time over the Internet. This function focuses on designing an effective and user-friendly process.
Finally, the management system of measurement stations, the administrator has the ability to dynamically create, modify and delete objects, points and information of each station on the GoogleMap. In this way the processing (update, delete, add) of points is easier.

The A.Q.M.E.I.S.  application has been developed using open source software tools like HTML, Javascript, PHP and MySQL. HTML is the language for the Internet Interface design. The goal for HTML was to create a platform-independent language for constructing hypertext documents to communicate multimedia information easily over the Internet \cite {web2008}.  Javascript is a client-side scripting language that provides powerful extensions to the HTML used to compose web pages and is mainly utilised  for checking and validating web form input values to make sure that no invalid data are submitted to the server  \cite {Java1}. PHP is the most popular server-side programming language for use on Web servers.  Any PHP code in a requested file is executed by the PHP runtime, usually to create dynamic web page content. It can also be used for command-line scripting and client-side GUI applications. PHP is a cross platform programming language and  can be used with many relational database management systems (RDBMS)\cite {php1}. MySql is a high-performance, multi-thread, multi-user RDBMS  and is built around a client-server architecture.  Also MySQL uses the Structured Query Language standard giving a great deal of control over this relational database system  \cite  {mysql}. Finally, Apache server is responsible for expecting requests from various programs - users and then serve the pages, according to the standards set by the protocol HTTP (Hypertext Transfer Protocol) \cite {apache}.
 




\section{User Interface}
In this section the user interface and functions of this application are described. There are three (3) levels of user access (groups of users). On the first level the user has the ability to be informed in real time about the weather conditions, the air pollution and air pollution indices in an area of interest using Google Map.The second level is for authorized users only, who can be informed analytically through reports about measurements of a specific time period. The third level is for the administrator, who has access to all information and who also inserts, updates or deletes data from the database. The administrator can also interfere dynamically and manage all the information of the GoogleMap.

\subsection{Online Web Station Reports}
 
The 'Online Web Station Reports' is a new online web feature which offers  to the approved members of the application to monitor, analyze, check and process the measurements using statistics of each station in real time. Furthermore, the ability of pumping previous measurements is given; a function that did not exist in the web application until a short time ago.
	The login  is achieved through the use of a special personal password that is given to the members by the support group of the web application. The users input  the password and after validation, they can perform a number of available functions in a safe and user-friendly online web environment.
More specifically the feature offers the following functions to its members:
Presentation of daily, weekly, monthly and according to the user's choice values either for a chosen station of all its measured data or for specifically chosen measured parameters (sensors), with simultaneous calculation and presentation of maximum, minimum, average values, sum, number and percentage of measurements in a table or the ability of output of data in MS Excel. The functions of the new web features are described in detail .
There are four categories of Online Web station reports, i.e. daily, weekly, monthly and custom. Each report is displayed in three parts (forms). In the first form, by choosing a station the image is displayed as well as various information about the specific station (Fig. 1).

\begin{figure}[!t]
	\centering
	\includegraphics[width=0.5\textwidth]{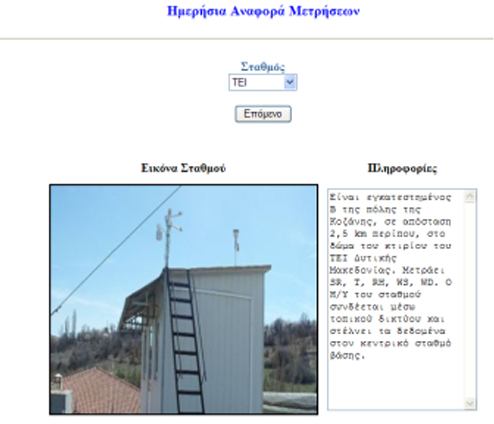}
	\caption{}
\end{figure}

If the user does not choose a station, an error report appears. By clicking on 'Next', the second form of the report appears in which the user can choose which measure fields to be shown,as well as  the measure time interval (5min or 60min) and the specific date that those measurements were taken (Fig. 2). 
\begin{figure}[!h]
	\centering
	\includegraphics[width=0.5\textwidth]{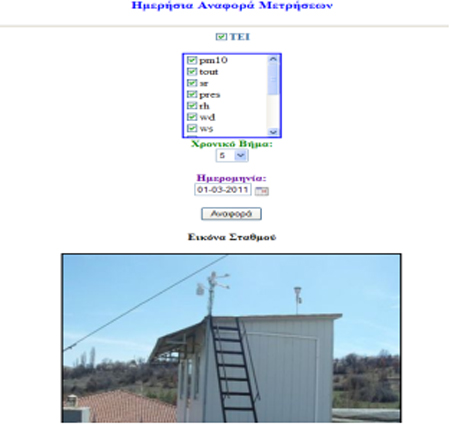}
	\caption{}
\end{figure}

The dates differentiate according to the report category that the user will choose; more specifically, there are:
a)	Daily report: the current date appears.
b)	Weekly: the first day of the current week is set as the starting date.
c)	Monthly: the first day of the current month is set as the starting date.
d)	Periodical: the current dates are set as the dates (From - To) with the ability of changing the spaces (From - To) by the user

If the user does not choose any measure fields or chooses date in which there are no figures reported, then the system will display an error message. By clicking on 'Report' the algorithm moves to the last tab of the report where a table of contents appears in a dynamic way with information about the hour, the measure fields, measurement results as well as other statistical data (Fig. 3).

\begin{figure}[!h]
	\centering
	\includegraphics[width=0.5\textwidth]{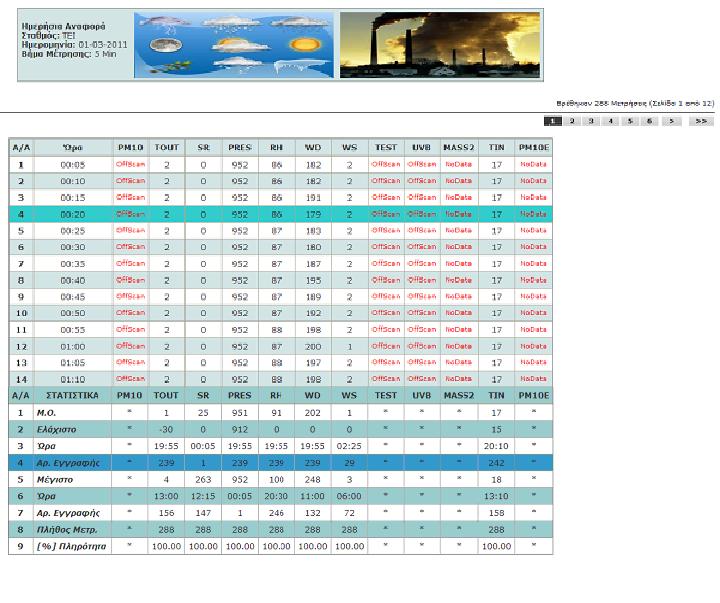}
     
 	\caption{}
\end{figure}

Also, the algorithm calculates and displays the number of measurements (e.g. '100 measurements were found'), the current page and the total number of pages (e.g. 'page 1 from 12'). Depending on the number of records, an equal number of pages is created. The application can display 25 measurements per page. Also the users can move to any page they wish so as to have access to any measurement of interest.  Every form of measurement also has a status field (numbers 0, 1 or 2). This table is used to check the validity of the measurements of a field. In this way, if there are no results for a specific date in one field, then the indication 'NO DATA' is displayed. All checks are made based on the status field. If,  however  the measurements in a field are wrong for a specific date  due to various factors, then the indication 'Offscan' is displayed. In a similar manner, checks rely on the status field. For every measure field in a specific moment the following statistics are taken into account a) average, b) minimum value, c) date and time that minimum value was found, d) the number of records of the minimum value, e) maximum value ,g) the date and  time of record of the maximum value, h) the total number of measurements, i) the \% percentage. By pressing 'Excel' the measurements of the station can be displayed on Excel form, which the user is able to open or save it  for later use (Fig. 4). 

\begin{figure}[!h]
	\centering
	\includegraphics[width=0.5\textwidth]{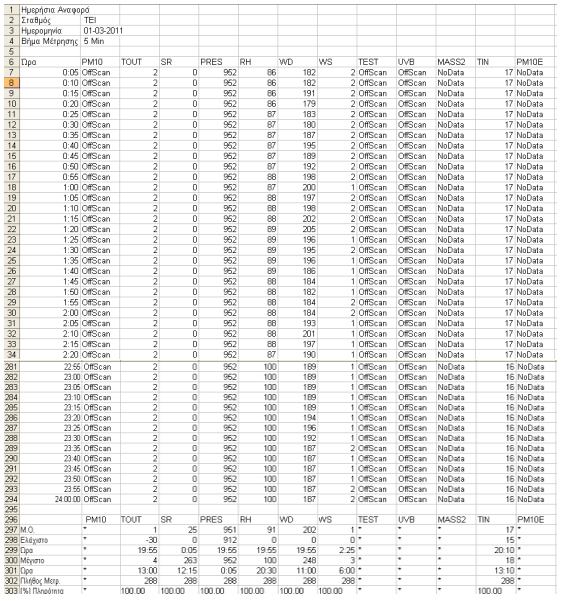}
 
\caption{}
\end{figure}

 \subsection{Manage Stations using GoogleMaps}

Another innovation of the web application is the dynamic management of the measurements from each station in a simple way using an online geographical web interface. The administrator using this specific feature (Management of the Measurement Stations using GoogleMaps) can insert, delete and modify easily and simply data having as a purpose the dynamic renewal on the GoogleMap. This gives the advantage to the administrator to use the specific feature as a platform of visualization of information without having to write on their own not even a line of code for this purpose. Moreover, an important element of the feature is the easy expansion and integration N (N = count) measurement stations on the interactive GoogleMap.

Station management is made through the interactive interface of GoogleMap; the administrator of this application can insert, delete and modify dynamically a certain point (station) in an area (according to geographical latitude and longitude). To insert a certain station in the map, the following actions are required: a) the insertion of Municipality of choice: the user chooses through a list the one which the station belongs to; b) the insertion of  the type of station of measurement: the user chooses if the station is meteorological or one that measures pollution or both. All the data is stored in the application MySQL database.  As a last step, the administrator sets the name of the station, the longitude and latitude, municipality, address, description, type of station and the image of the station. (Fig. 5). Next, all information are stored into the database and are retrieved from there to be displayed dynamically (both the points and information) on the GoogleMap.  

On the map users can see meteorological information as well as information about pollution from various stations and areas. For every station a previous and current index of pollution appears (in a scale 1-10) with an appropriate corresponding colour scale. By clicking on each point of the station the information (i.e. Online measurements, air pollution indices for the previous and current day, general information about the station) is displayed. The user may also activate or deactivate one or more points on the map \cite {TriantSkordas}.

To achieve the dynamic update of the measurement stations on the GoogleMap, the file airlab\_markers.php is called. It is responsible for the creation and update of the XML file. More specifically, the data of the  application, i.e. the name and the measurements of the station, their geographical position where they belong, the general information with the representative photograph of each station, even the representation symbol, are retrieved in XML structure, submitting the appropriate preset SQL question to the database, via the corresponding code of the php page.
The XML file has an element (root-top level element) and especially the '<markers> </markers>'. The remaining elements are nested to this. For the appropriate structure of the XML file, there is an additional code in the airlab\_markers.php file. All necessary checks about the validity of the data then take place. The algorithm was realized by PHP scripts and the specific feature that was developed, is supported by the Internet Explorer, Firefox, Opera and Google Chrome browsers.

\begin{figure}[!h]
	\centering
	\includegraphics[width=0.5\textwidth]{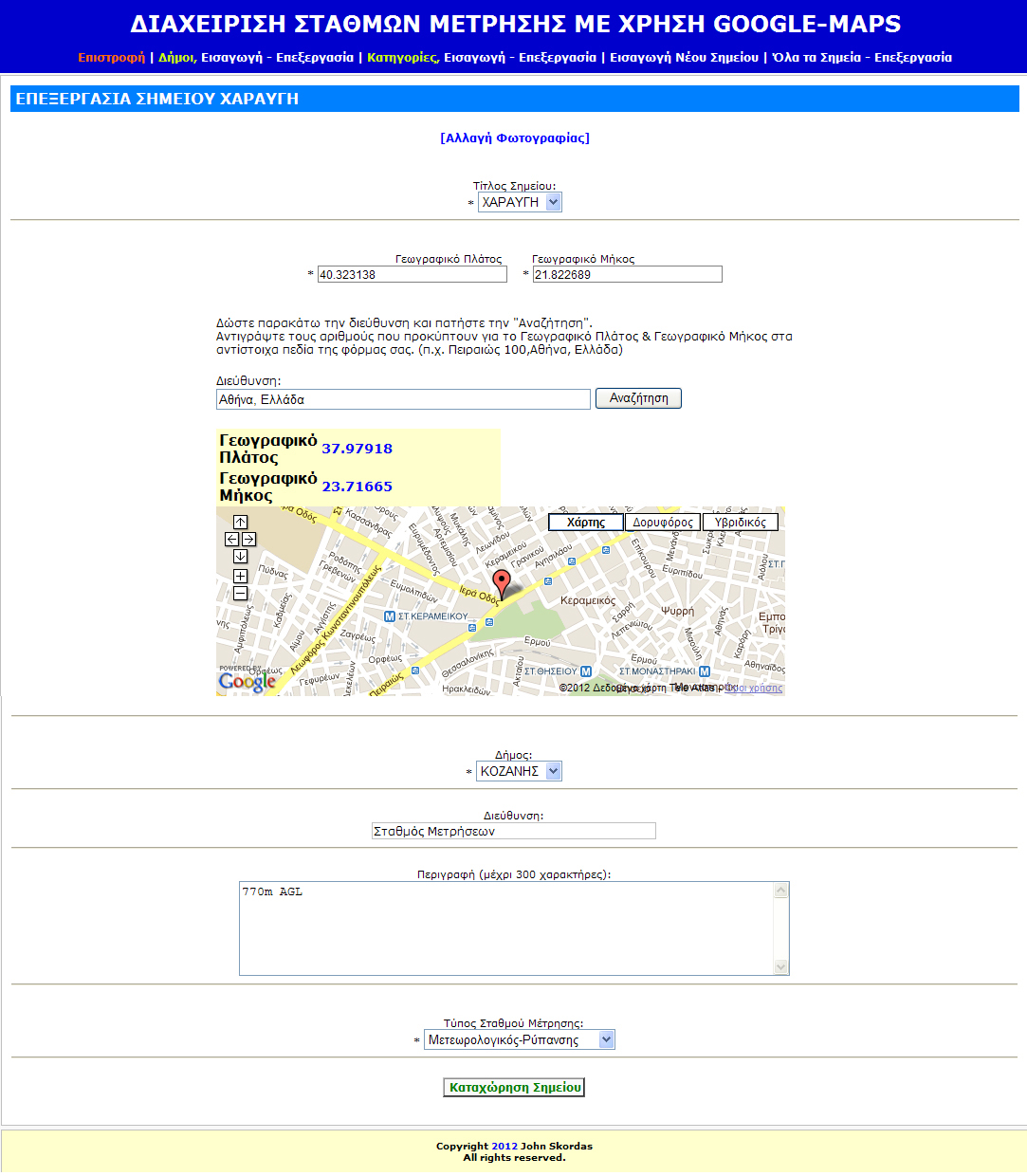}
\caption{}
\end{figure}

\subsection{Weather Forecast in West Macedonia with Dynamic Web Charts}
 
An additional new feature is the weather forecast with the aid of dynamic web charts, is committed to deliver  the most reliable, accurate weather information possible. It provides free, real-time and online weather information for the web users with the state-of-the-art technology monitors conditions and forecasts in the area of Western Macedonia in the next few days (Fig. 6).

The information is produced in a high-end server  in EAP / WMAQIS \cite {triantkrestou2011} and is read and stored in a database and  it appears in the internet with the form of dynamic web graphs (Fig.7).

 Meteorological parameters are temperature, humidity, wind speed, wind direction, accumulated precipitation, mixing height and total solar radiation. PHP scripts retrieves from the MySQL database the 24 average hourly values for each meteorological parameter (except for the accumulated precipitation, for which a total of 6 hours is taken into account) on each location in Western Macedonia. Next, the information is displayed with a graph. The user then can choose a location to see the weather forecast. By choosing 'History', the previous meteorological measurements and figures are displayed using graphs.
 
\begin{figure}[!h]
	\centering
	\includegraphics[width=0.5\textwidth]{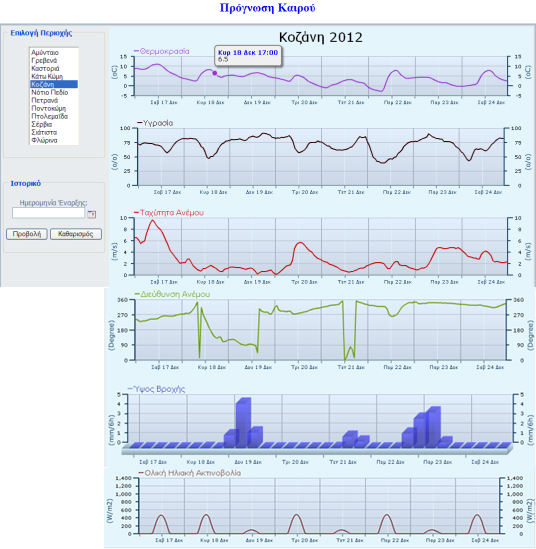}
 
	\caption{}
\end{figure}

\subsection{Air Pollution Forecast in West Macedonia}
Another important part of the A.Q.M.E.I.S. application is the atmospheric pollution forecast of the pm10 (particulate matter) concerning the next few days in Western Macedonia. (Fig. 8).
 Our application displays dynamically these regions in a map using images; according to pollution percentages in a certain region, the corresponding colour scale is represented denoting the levels of pollution. Choosing 'region', 'pollutant agent', 'source of emission' and 'date',  pollution for the previous and current dates, as well as the ones of the next three days are displayed . This part of the application uses javascript, while a very small part of the code was written in PHP (dates management).

The air pollution model produces image files  (xxx.jpg) in the hard disc of the server in which javascript searches and then displays. In cases where not enough environmental data exist for a certain date, an image appears entitled 'Pollution Image Display Unavailable'. The necessary validation integrity checks of the dates are also made (from-to). Finally choosing 'history' the user can see older images of pollution rates in a certain region of Western Macedonia.

\begin{figure}[!h]
	\centering
	\includegraphics[width=0.5\textwidth]{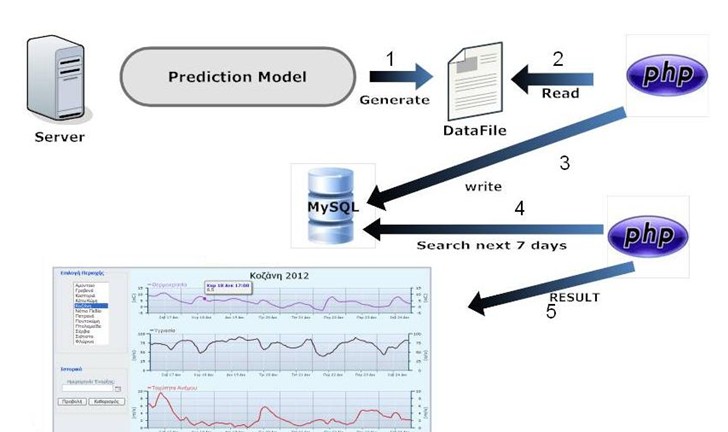}
	\caption{}
\end{figure}

Choosing 'movement', a javascript algorithm is executed animating the  pollution illustrations in the area of interest. (Fig 9).
 
\begin{figure}[!h]
	\centering
	\includegraphics[width=0.5\textwidth]{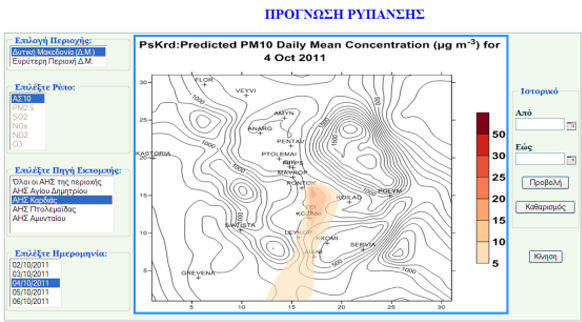}
	\caption{}
\end{figure}

 \begin{figure}[!h]
	\centering
	\includegraphics[width=0.5\textwidth]{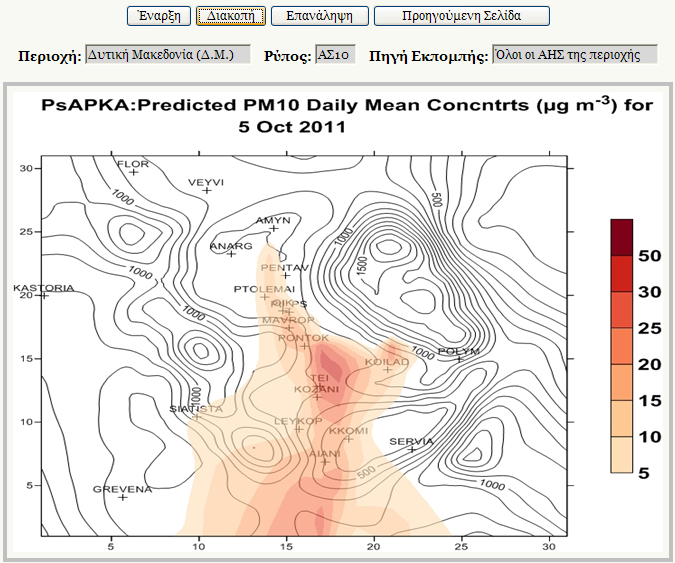}
	\caption{}
\end{figure}

\section{MySQL and Database Architecture}
MySql is a very fast and powerful database management system. Allows the user to store, search, sort and recall data efficiently. This application stores all information in the MySQL database, so that they can be retrieved  dynamically every time they are needed. The architecture of this database consists of a total of 23 tables. 8 tables (s001t05, s001t60, s002t05, s002t60, s003t05, s003t60, s004t05, s004t60) include measurements collected from the stations; they are used in reports and in the dynamic system for monitoring the air pollution, via an interactive chart. 's00x' refers to station x and t05 or t60 refer to a measure time interval (i.e. an average 5 or 60 minute measurement). The primary key is Date\_Time, while the rest of the fields (value1 up to value32)  store meteorological and environmental data.  Tables 'city\_info', 'points\_categories' and 'points' are used for the management of stations that use Google Map. In particular, 'city\_info' stores the town in which each point (station) is located, along with additional information. Fields include 'id' (main key), title, en\_title, lat, log . Table 'points\_categories' stores the station's category (fields: 'id' is the primary key, 'title' and 'en\_title') .  On the table 'points'  points are set up, along with additional information. These fields are 'id' (primary key), 'category', 'city', 'en\_city', 'address',  'title', 'description', 'lat', 'log', 'thumb', 'image' .  Finally 12 tables  (mamyntaio, mflorina, mgrevena, mkastoria, mkkomi, mkozani, mnpedio, mpetrana, mpontokomi, mptl, mservia, msiatista) which store weather forecast information are used.  Every table represents a certain location in Western Macedonia. Fields DATE , HOUR  have been set up as primary key denoting solely each record. The rest (WDIR, TEMP, RHUM, TEMPSCR, RHUMSCR, TSR, NETR, SENS, EVAP, WSTAR, ZMIX, USTAR, LSTAR, RAIN and SNOW) are meteorological parameters . Tables about pollution forecasts do not exist, all measurements are stored in the hard disc of the server as image files. A sample table of weather forecast has the following form : 
mkozani (DATE, HOUR, WDIR, TEMP, RHUM, TEMPSCR, RHUMSCR,  TSR,  NETR, SENS,  EVAP, WSTAR, ZMIX, USTAR, LSTAR, RAIN and SNOW.)

The proposed A.Q.M.E.I.S. application is part of a system - air quality monitoring network, which was developed in the Laboratory of Atmospheric Pollution and Environmental Physics of Technological Education Institute of Western Macedonia, to monitor the air  quality in Western Macedonia area, with industrial focus on the region of  Prolemais - Kozani basin. This system was co-financed by the TEIWM, Regional Operational Programm 2000 - 2006 Western Macedonia and recently by the municipality of Kozani.  The architecture of this system is constituted by five terminal stations, which collect environmental information, the central station and a web server. Different technologies (ADSL, GPRS, ETHERNET) are used to transfer the data to the central station. The  data  are sent every half an hour to  the main station which collects the complete set of data and transfers them to the web server every sixty minutes, where under the    application proposed in this paper provides meteorological, environmental, weather and air pollution forecast data in West Macedonia area. Further details on the design of the above mentioned air quality monitoring  network can be obtained from \cite {Triant2004},\cite  {TriantSkordas}, \cite {Skordas_Fragulis_Triant2011}.

\section{Conclusion}

 An operational monitoring, as well as high resolution local-scale meteorological and air quality forecasting information system(A.Q.M.E.I.S.) for Western Macedonia, Hellas, has been developed and is operated by the Laboratory of Atmospheric Pollution and Environmental Physics / TEI Western Macedonia. In this paper the novelty of information system  is presented, in a dynamic, easily accessible and user-friendly manner. The application is developed using state of the art web technologies (Ajax, Google maps etc) and under the philosophy of the open source software that gives the ability to users/authors to update/enrich the code so that their augmentative needs are met.

\begin{thebibliography}{1}

\bibitem {Triant2004} Design of a web\-based information system for ambient air quality data in west Macedonia, Greece, A.G. Triantafyllou, V. Evagelopoulos, E.S. Kiros, C. Diamantopoulos, 7th Panhellenic (International) Conference of Meteorology, Climatology and Atmospheric Physics, Nicosia 28-30 September, 2004.
\bibitem {Council1}Council Directive 90/313/EEC of 7 June 1990 on the freedom of access to information on the environment. 
\bibitem	{Council2}Council Directive 92/72/EEC of 21 September 1992 on air pollution by ozone.
\bibitem	{Council3}Council Directive 1999/30/EC of 22 April 1999 relating to limit values for sulphur dioxide, nitrogen dioxide and oxides of nitrogen, particulate matter and lead in ambient air.
\bibitem	{Council4}Council Directive 96/62/EC of 27 September 1996 on ambient air quality assessment and management. 
\bibitem {Council5}Council Directive 90/313/ΕEC in expanding the level of access in information

\bibitem {Karatzas} Karatzas K. and J. Kukkonen, 2009,  COST Action ES0602 Workshop Proceedings “Quality of life information services towards a sustainable society for the atmospheric environment” ed. K. Karatzas and J. Kukkonen. 
\bibitem {Schimak}Schimak G., 2003 : Environmental data management and monitoring system UWEDAT. Environmental Modelling and Software 18, 573-580.
\bibitem {Xuan} Xuan Zhu, Allan P. Dale, 2001 : JavaAHP : a web\-based decision analysis tool for natural resource and environmental management, Environmental Modelling and Software 16, 251-262.
\bibitem {Triant_book} Atmospheric Pollution - Atmospheric Boundary Layer, Advances in Measurements Technologies, A.G. Triantafyllou, Ed. TEI of Western Macedonia, Kozani 2004, pp.256.

\bibitem {anakoinosi} http://www.airlab.edu.gr/arlb/Anakoinwseis/6\_dt02\_zwntanes\_metrhseis.pdf.
\bibitem {triantEvazoras2006}A.G. Triantafyllou, V. Evagelopoulos, S. Zoras, Design of a web\-based information system for ambient air quality data,Journal of Environmental Management, 80(3), 230-236 (2006).
\bibitem {Comeap} http://comeap.org.uk/
 
\bibitem {TriantSkordas} A.G. Triantafyllou, J. Skordas, Ch.Diamantopoulos, E. Topalis, "The new dynamic air quality information systemand its application in West Macedonia, Greece", p.4, 4th Macedonian Enviromental Conference, Thessaloniki, Greece, March 2011. 

\bibitem {Skordas_Fragulis_Triant2011} Ioannis Skordas, George F. Fragulis, Athanasios G. Triantafyllou, "e\-AirQuality: A Dynamic Web Application for Evaluating the Air Quality Index for the city of Kozani, Hellas", PCI 2011 15th, Panhellenic Conference on Informatics, 30 September - 02 October  2011, Kastoria, Greece.
\bibitem {airlab}	http://www.airlab.edu.gr/
 
\bibitem {web2008} Beginning Web Programming with HTML, XHTML, and CSS, Jon Duckett, Wiley Publishing, 2008.
\bibitem {php1}Core PHP Programming 3rd Edition, Leon Atkinson with Zeev Suraski, Pearson Education, Inc,  Publishing as Prentice Hall Professional Technical Reference, 2004 
\bibitem {mysql} MySQL: The complete Reference, Vikram Vaswani, McGraw-Hill \- Brandon A. Nordin, 2004.
\bibitem {apache} Apache: The Definitive Quide 3rd Edition, Ben Laurie and Peter Laurie, O'Reilly, 2003. 
\bibitem {Java1} http://www.w3schools.com/js/
\bibitem {triantkrestou2011}Triantafyllou A.G., Krestou A., Hurley P. and Thatcher M., "An operational  high resolution local\-scale meteorological and air quality forecasting system   for Western Macedonia, Greece: Some first results", Proceedings of the 12th International Conference on Environmental Science and Technology (CEST2011),8\- 10 September 2011, pp.A\-1904:1911


\end{thebibliography}
\end{document}